\documentclass[modern]{aastex63}

\usepackage{natbib,graphics,graphicx,mhchem,amssymb,lineno} 
\bibpunct{(}{)}{;}{a}{,}{,}

\newcommand{\GG}[1]{}
\usepackage[symbol]{footmisc}

\begin{document}

\title{Observations of Titan's Stratosphere During Northern Summer:
  Temperatures, CH$_3$CN and CH$_3$D Abundances}

\correspondingauthor{Alexander E. Thelen}
\email{athelen@caltech.edu} 

\renewcommand{\thefootnote}{\fnsymbol{footnote}}
\setcounter{footnote}{1}

\author{Alexander E. Thelen}
\affiliation{Division of Geological and Planetary Sciences, California
  Institute of Technology, Pasadena, CA 91125, USA}
\affiliation{Solar System Exploration Division, NASA Goddard Space Flight Center, Greenbelt, MD 20771, USA}

\author{Conor A. Nixon}
\affiliation{Solar System Exploration Division, NASA Goddard Space Flight Center, Greenbelt, MD 20771, USA}

\author{Martin A. Cordiner}
\affiliation{Solar System Exploration Division, NASA Goddard Space
  Flight Center, Greenbelt, MD 20771, USA}
\affiliation{Department of Physics, Catholic University of America, Washington,
DC 20064, USA}

\author{Emmanuel Lellouch}
\affiliation{LESIA, Observatoire de Paris, Universit{\'e} PSL,
  Sorbonne Universit{\'e}, Universit{\'e} Paris
  Cit{\'e}, CNRS, 92195 Meudon, France}

\author{Sandrine Vinatier}
\affiliation{LESIA, Observatoire de Paris, Universit{\'e} PSL,
  Sorbonne Universit{\'e}, Universit{\'e} Paris
  Cit{\'e}, CNRS, 92195 Meudon, France}

\author{Nicholas A. Teanby}
\affiliation{School of Earth Sciences, University of Bristol, Bristol, UK}

\author{Bryan Butler} 
\affiliation{National Radio Astronomy Observatory, Socorro, NM 87801, USA}

\author{Steven B. Charnley}
\affiliation{Solar System Exploration Division, NASA Goddard Space
  Flight Center, Greenbelt, MD 20771, USA}

\author{Richard G. Cosentino}
\affiliation{Space Telescope Science Institute, Baltimore, MD 21218, USA}

\author{Katherine de Kleer}
\affiliation{Division of Geological and Planetary Sciences, California
  Institute of Technology, Pasadena, CA 91125, USA}

\author{Patrick G.J. Irwin}
\affiliation{Atmospheric, Oceanic and Planetary Physics, Clarendon
  Laboratory, Oxford OX1 3PU, UK}

\author{Mark A. Gurwell}
\affiliation{Center for Astrophysics $|$ Harvard $\&$ Smithsonian,
  Cambridge, MA 02138, USA}

\author{Zbigniew Kisiel}
\affiliation{Institute of Physics, Polish Academy of Sciences,
  Al. Lotnik{\'o}w 32/46, 02-668 Warszawa, Poland}

\author{Raphael Moreno}
\affiliation{LESIA, Observatoire de Paris, Universit{\'e} PSL,
  Sorbonne Universit{\'e}, Universit{\'e} Paris
  Cit{\'e}, CNRS, 92195 Meudon, France}

\begin{abstract}
Titan's atmospheric composition and dynamical state have
previously been studied over numerous epochs by both ground- and space-based
facilities. 
However, stratospheric measurements
remain sparse during Titan's northern summer and fall. The lack of
seasonal symmetry in observations of Titan's temperature field and chemical
abundances raises questions about the nature of the
middle atmosphere's meridional circulation and
evolution over Titan's 29-yr seasonal cycle that can only be answered
through long-term monitoring campaigns.
Here, we present maps of Titan's stratospheric temperature, acetonitrile
(or methyl cyanide; CH$_3$CN), and monodeuterated methane (CH$_3$D)
abundances following Titan's northern summer solstice obtained with
Band 9 ($\sim0.43$ mm) ALMA observations. We find that
increasing temperatures towards high-southern latitudes, currently
in winter, resemble those observed during Titan's northern
winter by the Cassini mission. Acetonitrile abundances have changed significantly since previous
(sub)millimeter observations, and we find that the species is now highly concentrated at
high-southern latitudes. The stratospheric CH$_3$D content is found to
range between 4--8 ppm in these observations, and we infer the CH$_4$ abundance to vary between
$\sim0.9-1.6\%$ through conversion with previously measured D/H
values. A global value of CH$_4=1.15\%$ was retrieved, lending further
evidence to the temporal and spatial variability of Titan's
stratospheric methane when compared with previous measurements. Additional observations are required to
determine the cause and magnitude of stratospheric enhancements in
methane during these poorly understood seasons on Titan. 
  
\end{abstract}

\section{Introduction} \label{sec:intro}
\renewcommand{\thefootnote}{\arabic{footnote}}
\setcounter{footnote}{0}
Titan's atmosphere contains relatively high amounts of methane
(CH$_4$) that, when dissociated along with molecular nitrogen (N$_2$),
forms an expansive network of trace chemical species (see, for example, the
reviews by \citealp{bezard_14a}, \citealp{horst_17}, and \citealp{nixon_24}). The distribution
of these nitriles and organics both influences and is influenced by Titan's global
circulation \citep{hourdin_95, rannou_04, newman_11,
    lora_15, lebonnois_14, lombardo_23a}, which changes with Titan's long ($\sim7$ yr) seasons during
which longer-lived photochemical products accumulate at the
shrouded winter pole (e.g., \citealp{teanby_08b, teanby_17}). These chemical products also impact Titan's
atmospheric energy budget through the absorption and radiation
of infrared and UV photons \citep{yelle_91, tomasko_08b, bezard_18}. Thus the temporal and spatial distribution of
Titan's trace species, and the effects of their radiative exchange with
the atmosphere, requires investigation of the middle
atmosphere (stratosphere and mesosphere; $\sim100-500$ km) throughout
Titan's 29.5 yr orbit. Great
strides in this formidable task have been enabled by the Cassini/Huygens
mission and the multitude of measurements presented in recent years on
the detection of trace molecules and isotopic ratios \citep{bezard_07,
  vinatier_07b, nixon_08a, jennings_08,
  coustenis_08, jennings_09, jolly_10, 
  nixon_12, nixon_13, jolly_15},
and seasonal variability in temperature, winds, and composition
\citep{coustenis_07, vinatier_07a, teanby_08a, teanby_08b,
  achterberg_08, coustenis_10, vinatier_10, teanby_10a, achterberg_11,
  cottini_12, teanby_12, 
  vinatier_15, coustenis_16, teanby_17, sylvestre_18, coustenis_18, teanby_19,
  coustenis_20, sharkey_20, vinatier_20, mathe_20, sharkey_21, achterberg_23}. These studies form the foundation for
continued monitoring of Titan's atmospheric state during portions of
Titan's year beyond the Cassini mission (2004--2017; L$_S\sim300-90^\circ$), provoking questions
and an impetus for further investigations of novel chemical pathways, exogenic
energy sources and processes, and the
origin and resupply of its crucial methane reservoir
\citep{tobie_06, lunine_08, lellouch_14, glein_15, davies_16,
  nixon_18, miller_19}.

Following the end of the Cassini mission near Titan's
  northern summer solstice, ground-based investigations may provide
  insights into changes occurring in the stratosphere and lower
  mesosphere as influenced by the heretofore unmonitored
  seasonal changes. Chemical species with both relatively short (\textless1 yr) and long 
  (e.g., 10s to 100s of yr) photochemical lifetimes can be used as a
  probe of global scale atmospheric dynamics driven by Titan's
  pole-to-pole circulation cell (and near the equinoxes, two
  equator-to-pole cells). Species such as hydrogen cyanide (HCN) and
  CH$_3$CN, with stratospheric lifetimes of
 10s of years \citep{yung_84, wilson_04, krasnopolsky_09,
   krasnopolsky_14, dobrijevic_14, loison_15, willacy_16, vuitton_19},
  show distributions with lingering enhancements over Titan's winter
  pole until the spring. This is largely due to the accumulation of
  trace species from the subsiding branch of the global circulation
  cell during winter, which are then confined within the boundary of the circumpolar
  vortex at latitudes
  \textgreater60$^\circ$ and persist due to relatively slow or
  inefficient photochemical destruction \citep{coustenis_10, teanby_08a, teanby_09a,
    teanby_09b, teanby_10b, vinatier_15, vinatier_20, mathe_20, sharkey_20,
    sharkey_21, achterberg_23}. These
  distributions can be contrasted with gases of short photochemical
  lifetimes (e.g., cyano acetylene; HC$_3$N), which exhibit rapid
  enhancements over the fall pole shortly after equinox
  and are depleted at other latitudes
  \citep{thelen_19a, teanby_19, vinatier_15, cordiner_19,
    vinatier_20}.

 Additionally, there exists the curious case of Titan's stratospheric methane,
  which is thought to be uniformly mixed over seasonal timescales
  \citep{wilson_04, niemann_05}, but has been
  shown to potentially be modified by convective injection from
  localized regions in the troposphere \citep{lellouch_14}. Though Titan's methane distribution was
thought to be relatively constant above the troposphere and may not
show substantial variability with latitude, altitude, and time (often
defined using the Huygens measurements presented in \citealt{niemann_10}),
\citet{lellouch_14} inferred that the CH$_4$ mixing ratio varied from $\sim1$--$1.5\%$
between Titan's tropical ($\sim0$--$20^\circ$) and temperate ($\sim30$--$40^\circ$) latitudes through measurements
with the Cassini Composite Infrared Spectrometer (CIRS)
instrument. Subsequent analysis of 4 Cassini occultations 
with the Visual Infrared Mapping Spectrometer (VIMS), initially by
\citet{maltagliati_15} and reanalyzed in \citet{rannou_21} and
\citet{rannou_22}, 
resulted in stratospheric methane measurements closer to $\sim1.1\%$, with localized vertical and
latitudinal enhancements. A re-analysis of spectra recorded by the Descent Imager
and Spectral Radiometer (DISR) instrument during the descent of the
Huygens probe also suggests that the CH$_4$ mixing ratio decreases with
altitude in Titan's stratosphere, reaching a value of $\sim1\%$
at altitudes above 110 km \citep{rey_18}. Together, the measurement of CH$_4$ and various trace
gases allow for the inference of global, and potentially local,
circulation and meteorological events that affect the composition and
dynamics of the stratosphere, which in turn affect the energy budget
of the upper atmosphere and precipitation from the troposphere onto the surface
\citep{yelle_91, hourdin_95, rannou_04, rannou_06, crespin_08, tomasko_08b,
  mitchell_09, mitchell_12, lebonnois_12, lebonnois_14, lora_15, lombardo_23a}.

In recent years, far-IR and (sub)millimeter observations from facilities such as
the Institut de Radioastronomie Millim{\'e}trique (IRAM) 30-m
telescope and interferometer,
Submillimeter Array (SMA), the Atacama Large
Millimeter/submillimeter Array (ALMA), and the Herschel Space
Telescope have provided the means by which
to complement and continue the investigation of Titan's substantial,
complex atmosphere. Early ground- and space-based (sub)mm observations
of Titan allowed for
the confirmation of H$_2$O and HC$_3$N following their detections with
the Infrared Space Observatory (ISO; \citealp{coustenis_98}) and \textit{Voyager 1}
spacecraft \citep{kunde_81}, the first detections of
HNC and acetonitrile (or methyl cyanide; CH$_3$CN), and the derivation of vertical abundance and
temperature profiles 
\citep{bezard_92, bezard_93, marten_02, gurwell_04, courtin_11, rengel_11, moreno_11,
  moreno_12b, rengel_14, bauduin_18, rengel_22}. Highly resolved spectroscopic
measurements provided the means by which to derive Titan's wind speeds
between the upper stratosphere and the thermosphere
through Doppler shifts \citep{moreno_05, lellouch_19, cordiner_20, light_24}. Through the measurement of rotational transitions
of thermal tracers such as carbon monoxide (CO) and HCN, ALMA has
recently allowed for the derivation of Titan's vertical
temperature profiles throughout the stratosphere, mesosphere, and into
the lower thermosphere \citep{serigano_16, thelen_18, lellouch_19,
  thelen_22}. Additionally, its high spectral resolution and extensive
frequency coverage enable resolved emission line profiles of many
organic molecules -- some of which were unable to be detected by Cassini
in the infrared, including C$_2$H$_5$CN, C$_2$H$_3$CN, and others
\citep{cordiner_15, palmer_17, cordiner_19, thelen_19a, thelen_20, nixon_20}. Finally, ALMA has
provided the means by which to monitor Titan's atmospheric CH$_4$
content in the stratosphere by way of monodeuterated methane
(CH$_3$D; \citealp{thelen_19b}), which has previously been measured in
the IR through ground- and space-based facilities and used to derive
Titan's deuterium-to-hydrogen ratio (D/H). Early measurements from
Voyager and ground-based IR observatories are discussed in
\citet{penteado_05}, and references therein, while the values derived
from Cassini data are detailed in \citet{nixon_12}. The distribution and modification of
these molecular abundance profiles can provide global snapshots of the
dynamical state of the atmosphere, while the derived temperature
profiles and winds measured from Doppler shifts helps to complete the picture of the seasonal evolution
of Titan's atmosphere \citep{horst_17, nixon_18, teanby_19}.

Here, we present the analysis of ALMA data acquired in 2022
June ($L_S\approx146^\circ$, during the transition of Titan's
  northern hemisphere from summer into fall), designed to investigate both the long-term evolution of Titan's
temperature and chemical abundances. In particular,
  the measurement of rotational CH$_3$D transitions allows for a comparison
  to recent Cassini studies showing a CH$_4$ distribution influenced by
  complex tropospheric and stratospheric interplay; further, ALMA
  allows for the vertical and spatial distribution of CH$_3$CN to be investigated
  for the first time, providing an additional tracer of seasonal
  dynamics \citep{thelen_19a}. The results of CH$_3$CN mapping can be
  compared to HCN and other long-lived chemicals that were measured by
  Cassini (e.g., CO$_2$, C$_2$H$_6$) throughout Titan's year. These
  measurements, along with Titan's temperature field, are investigated here
during a period with very limited prior observational coverage. The
observational details are described in Section \ref{sec:obs}, followed
by the description of the radiative transfer modeling employed to
derive temperature and abundance information in Section \ref{sec:rad}. The
resulting atmospheric retrievals and the discussion, comparisons to prior measurements, and
implications are presented in Section \ref{sec:dis}. A summary
of our conclusions 
is provided in Section \ref{sec:conc}.

\section{Observations} \label{sec:obs}
The primary array of the ALMA facility is comprised of 50 12-m antennas located on the Chajnantor plateau
in the Atacama Desert, Chile. Titan was observed by
44--48 antennas on 2022 June 18 and 29 in ALMA Band 9
($602-720$ GHz; $\sim0.4-0.5$ mm), simultaneously targeting the
rotational transitions of CO ($J=6$--5), HCN ($J=8$--7), CH$_3$CN
($J=38$--37), and the CH$_3$D $J=3$--2 triplet located
 between $690-710$ GHz for ALMA Project
 $\#$2021.1.01388.S. The observed rotational transitions are detailed
 in Table \ref{tab:spec}. The targeted spectral windows were set to
 resolutions of 488
 kHz (HCN), 977 kHz (CH$_3$CN, CH$_3$D), and 1129 kHz (CO, continuum
 window). Millimeter observations at high frequencies are
significantly impacted by the Earth's atmosphere directly above the
facility through the reduced atmospheric transmittance and the
precipitable water vapor content, which produces interferometric phase
artifacts that reduce the image coherence (analogous to optical
`seeing' effects). As such,
  the high frequency transitions of many known trace chemical species have yet
  to be observed on Titan (or beyond) through
  interferometry. While \citet{serigano_16} and
    \citet{thelen_19a} observed high frequency transitions of CO
    ($J=6$--5), HCN ($J=8$--7),
  and CH$_3$CN ($J=37$--36), the $J=3$--2 rotational CH$_3$D transitions have yet to be
  definitively observed in astrophysical environments (though the
  $J=2$--1 transitions were previously detected on Titan -- see \citealp{thelen_19b}).

\begin{deluxetable}{llll}
   \tablecaption{Observed Spectral Transitions}
   \tablecolumns{4}
   \tablehead{Species & Transition$^a$ & Rest Freq. & $E_u$ \\
   & & (GHz) & (K) \\ [-2.75ex]}
  \startdata
  %Species  Transition Rest Freq.  E"
CO & $6\rightarrow5$ & 691.473 & 116.16 \\
CH$_3$D & $3_{2}\rightarrow2_{2}$ & 697.691 & 74.86 \\
 & $3_{1}\rightarrow2_{1}$ & 697.759 & 68.95 \\
 & $3_{0}\rightarrow2_{0}$ & 697.781 & 66.98 \\
CH$_3$CN & $38_{9}\rightarrow37_{9}$ & 697.209 & 1230.80 \\
 & $38_{8}\rightarrow37_{8}$ & 697.434 & 1109.86 \\
 & $38_{7}\rightarrow37_{7}$ & 697.632 & 1003.08 \\
 & $38_{6}\rightarrow37_{6}$ & 697.804 & 910.50 \\
 & $38_{5}\rightarrow37_{5}$ & 697.949 & 832.12 \\
 & $38_{4}\rightarrow37_{4}$ & 698.068 & 767.97 \\
 & $38_{3}\rightarrow37_{3}$ & 698.161 & 718.06 \\
 & $38_{2}\rightarrow37_{2}$ & 698.227 & 682.40 \\
 & $38_{1}\rightarrow37_{1}$ & 698.267 & 661.00 \\
 & $38_{0}\rightarrow37_{0}$ & 698.281 & 653.86 \\
  \enddata
   \footnotesize
   \tablecomments{Spectral line positions and upper level energies ($E_u$) are taken from the Cologne
Database for Molecular Spectroscopy\footnote{https://cdms.astro.uni-koeln.de/classic/entries/} (CDMS; \citealp{muller_01,
  muller_05, endres_16}). $^{a}$Rotational transitions are denoted as
     $Ju \rightarrow Jl$ or $Ju_{Kl} \rightarrow Jl_{Kl}$, where $u$
     and $l$ represent the upper and lower energy states,
     respectively, and $K$ represents the angular momentum quantum number.}
   \label{tab:spec}
 \end{deluxetable}
  
Fortunately, the relatively
low ($\sim0.3$--0.5 mm) precipitable water vapor measurements during
the observation dates resulted in low phase scatter, which facilitates observations with the
moderately extended ALMA array, which was set to configuration C-5,
with maximum antenna separations of up to $\sim1.4$
km. While the ALMA data were reduced with the standard pipeline
procedures provided by the Joint ALMA Observatory in the Common Astronomy
Software Applications (CASA; \citealp{jaeger_08}) ver. 6.2,
additional iterative self-calibration and imaging
procedures were performed so as to improve
the image coherence and signal-to-noise (S/N) ratio, executed similarly to those used for the Galilean Satellites and Giant
Planets \citep{de_pater_19, de_kleer_21a, camarca_23, thelen_24a}. First, the CH$_3$CN spectral
window and a relatively featureless spectral window from the
correlator upper sideband at $\sim712$ GHz were used to create a
single continuum image of Titan by combining data from both ALMA executions. We flagged out all strong spectral
line channels from atmospheric trace species and then concatenated the
remaining averaged continuum 
channels using the multi-frequency synthesis
settings in the CASA \texttt{tclean} algorithm. Using a
starting model of a flat or limb-darkened disk at the appropriate
brightness temperature of Titan's Band 9 continuum, we solved for
phase solutions as a function of time on this continuum image with successively
smaller solution intervals until phase scatter due to noisy antenna
baselines were sufficiently minimized; see the reviews in \citet{cornwell_99b},
\citet{butler_99}, \citet{brogan_18}, and ALMA Memo $\#$620\footnote{Richards
  et al., ALMA Memo
  $\#$620:https://library.nrao.edu/public/memos/alma/main/memo620.pdf}
for further details on self-calibration of bright,
compact sources such as planetary disks. Once the continuum phase
distribution was on order $\pm10\%$, we applied these phase solutions to the full
calibrated spectral line data and performed further image
deconvolution on the combined executions to obtain the
final spectral image cube. This round of imaging was
performed using the CASA \texttt{tclean} task with the
  H{\"o}gbom algorithm, natural antenna weighting, and $0.01''$
  square pixels, resulting in a final angular resolution (represented
  by the ALMA beam) of
  $0.172\times0.148''$ ($\sim1000$ km on Titan at the time of
  observation) with a position angle of $71.49^\circ$. Compared to Titan's angular size of
$\sim1.0''$ (including its 2575 km solid-body radius and atmosphere up
to 1200 km) during the time of observation, this resolution
  allowed us to observe localized emission from a number of distinct 
latitude regions ranging from $\sim77^\circ$S to $90^\circ$N.

\begin{figure}
  \centering
  \includegraphics[scale=0.7]{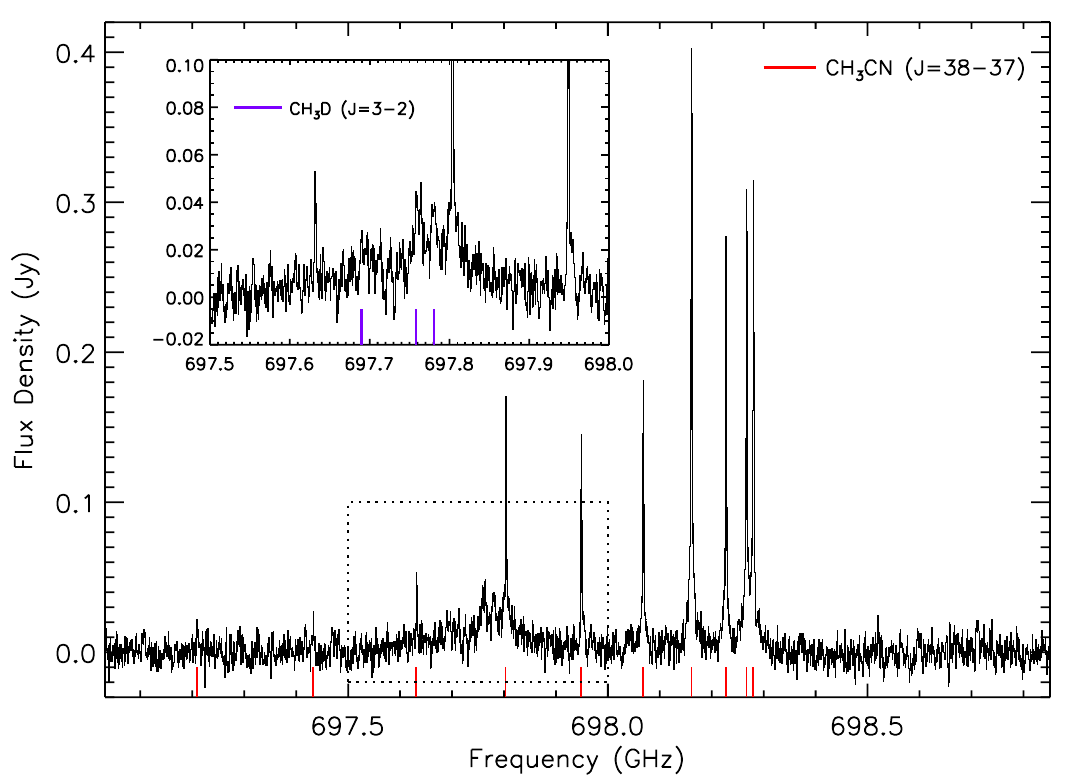}
  \caption{Limb-averaged Titan spectrum, created by averaging 9
    separate 20$^\circ$ latitude bins
    from both eastern and western
    hemispheres following the combination of both ALMA
    executions during 2022 June. The spectral resolution was 977 kHz. Spectra were extracted from $\sim150$
    km above the surface. Data are shown following continuum
    subtraction. Transitions of the CH$_3$CN ($J=38$--37) band from the
    CDMS catalogue, Doppler-shifted to Titan's rest velocity, are
    shown in red for reference; the line marker amplitudes are
    arbitrary. The dotted box denotes the spectral range of the
    figure inset, which focuses on the
    CH$_3$D ($J=3$--2) triplet. The CH$_3$D transitions from the CDMS
    catalogue are
    again denoted in vertical purple lines in the inset. As a result
    of the continuum subtraction, the extent of the pressure-broadened
    CH$_3$D band is evident, in comparison to the relatively narrow
    CH$_3$CN emission lines.}
  \label{fig:spec}
\end{figure}

\begin{figure}
  \centering
  \includegraphics[scale=0.6]{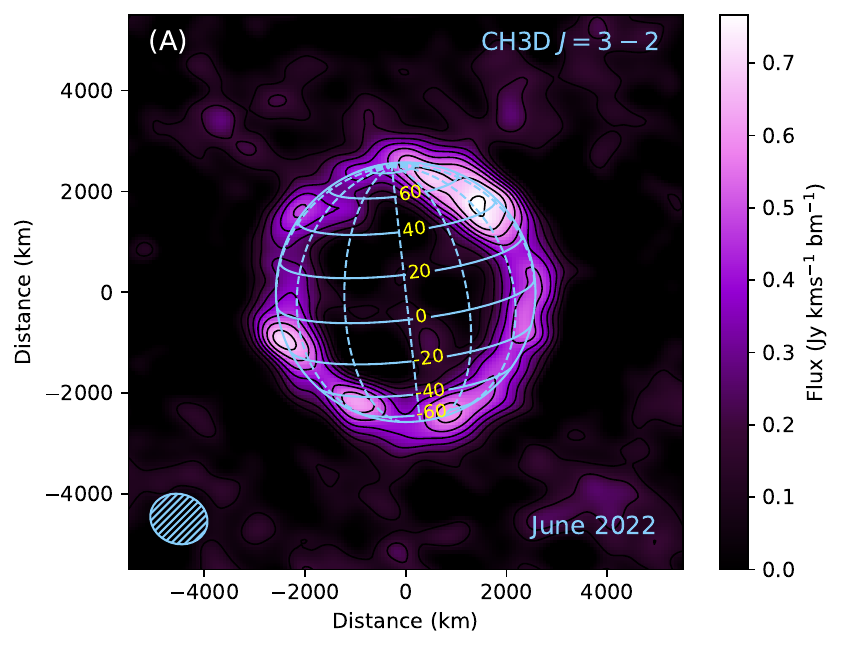}
  \includegraphics[scale=0.6]{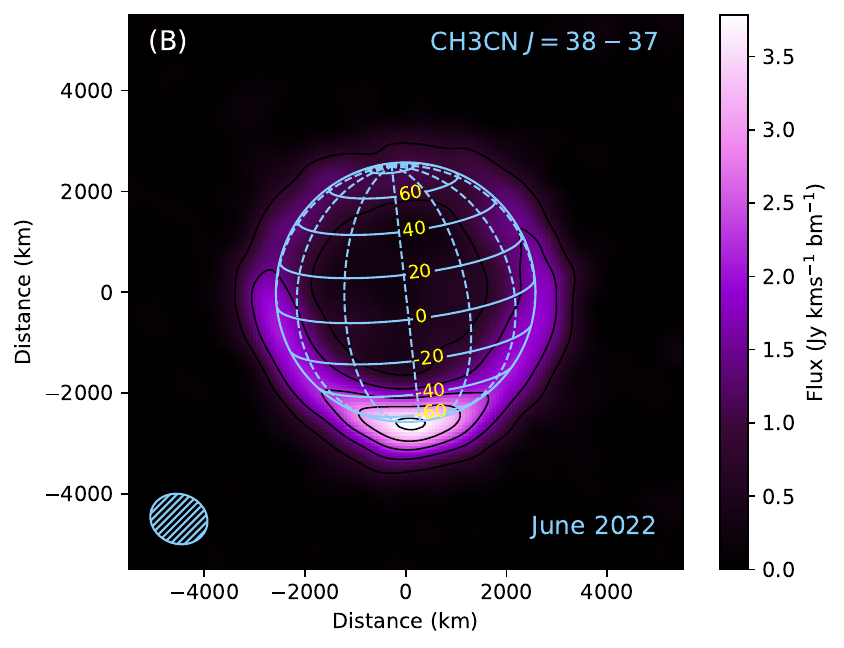}
  \includegraphics[scale=0.6]{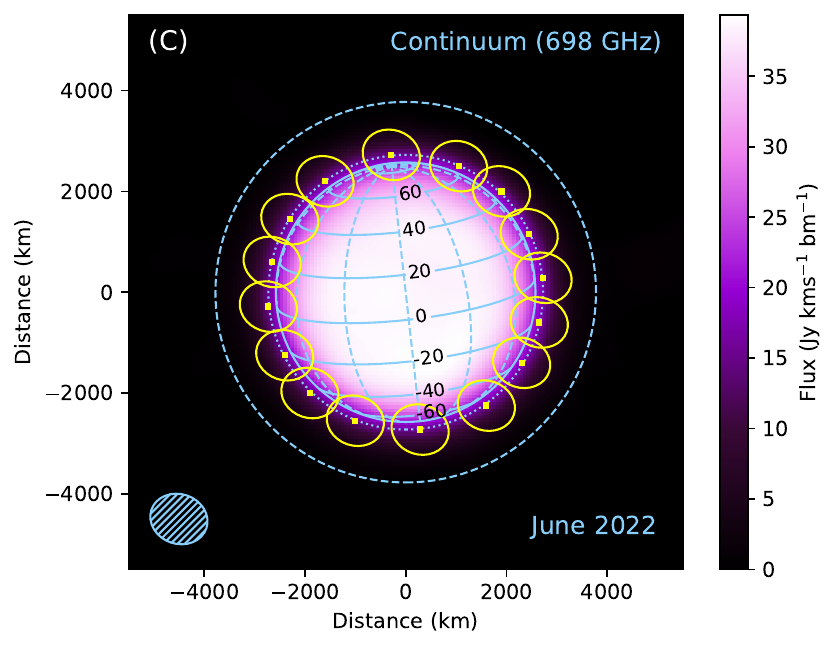}
  \caption{Integrated emission (or `moment 0') maps of Titan's CH$_3$D
    (A), CH$_3$CN (B), and continuum (C) spectra. Panels (A) and (B) correspond to the
    emission integrated over the spectral transitions shown in Fig. \ref{fig:spec}. The synthesized
    ALMA beam size (FWHM of the ALMA PSF) is shown as the hatched
    ellipse in the lower left; Titan's solid
    surface, latitude and longitudes are shown (solid and dashed blue lines). Contours are in increments of 1$\sigma$ (A) and
    10$\sigma$ (B). Spectra were extracted for
  radiative transfer modeling at $20^\circ$ latitude bins as
  displayed on the continuum emission map in panel C (yellow squares), and the
  corresponding area representing the ALMA beam foot-print used to generate synthetic
  spectra from these individual regions on Titan's limb are shown (yellow
  ellipses). Altitudes of 150 km and 1200 km above Titan's surface are
denoted by dotted and dashed blue circles, respectively, on the
continuum map.}
  \label{fig:maps}
\end{figure}

A latitudinally averaged spectrum extracted from $\sim150$ km above
Titan's solid surface, where the CH$_3$D emission is strong, is presented in Figure \ref{fig:spec},
showing the stronger CH$_3$CN $J=38$--37 transitions surrounding the broad CH$_3$D $J=3$--2 triplet. Emission
maps produced by the integration of channels across this spectral
range at every pixel are shown in Figure \ref{fig:maps} for the
CH$_3$D and CH$_3$CN rotational bands, as well as an image of the
continuum emission for reference. While the
emission from CH$_3$CN is localized at high-southern latitudes, the
CH$_3$D emission map possesses a significantly lower signal-to-noise ratio, preventing
substantial conclusions regarding variations in the CH$_3$D distribution to be drawn from imaging alone. As in previous
studies using ALMA to observe Titan, we extracted spectra at spatially
independent regions to determine the variability of vertical profiles
with latitude (see Figure \ref{fig:maps}C). The spatial resolution of these observations enabled
the analysis of spectra from approximate $20^{\circ}$ latitude bins, which were
averaged with the surrounding $5\times5$ pixel grid such as to improve the spectral
S/N.

\section{Radiative Transfer Modeling} \label{sec:rad}
Latitudinally averaged spectra were extracted from the regions denoted in
Figure \ref{fig:maps}C (yellow squares) for use with the
radiative transfer package NEMESIS: the Non-linear optimal Estimator
for MultivariatE spectral analySIS (\citealp{irwin_08}; NEMESIS is
publicly available
online\footnote{https://nemesiscode.github.io}). Wavenumber offsets on
order $1\times10^{-5}$ cm$^{-1}$ were added to spectra where necessary
to account for minor Doppler shifts due to Titan's wind field
\citep{lellouch_19, cordiner_20}; this is $\sim\frac{1}{3}$ of the ALMA
channel spacing. 30 line-of-sight emission angles
were calculated for each individual spectrum extraction location so as to
correctly model the emission originating from the corresponding ALMA beam-footprint
distributed around Titan's limb (Figure \ref{fig:maps}C, yellow
ellipses); see \citet{thelen_18} for additional details on the
construction of emission angles to represent the ALMA beam shape.

Spectral line frequencies, broadening and temperature dependence parameters,
and partition function coefficients were taken from the CDMS and HITRAN database\footnote{https://hitran.org}
\citep{rothman_05, rothman_13, gordon_17, gordon_22} where
available. The N$_2$-broadening parameters for CH$_3$CN were taken
from \citet{dudaryonok_15b}, as discussed in \citet{thelen_19a}. Collisionally-induced absorption of 
N$_2$, CH$_4$, and H$_2$ pairs were parameterized as in previous
studies of Titan with ALMA and Cassini/CIRS \citep{borysow_86a,
  borysow_86b, borysow_86c, borysow_87, borysow_91,
  borysow_93}. Vertical profiles of these gases were used from
Huygens Gas Chromatograph Mass Spectrometer (GCMS) measurements
\citep{niemann_05, niemann_10}.

Following spectral extraction and conversion, NEMESIS forward models
were generated for spectral regions containing only continuum emission
(formed from the above absorption pairs and thermal radiation) to
determine constant, multiplicative scaling factors to apply to the spectra due
to flux calibration uncertainties; ALMA flux calibration uncertainties at
high frequencies can be as high as $20\%$\footnote{See the ALMA Cycle
  8-2021 Proposers Guide: https://almascience.eso.org/documents-and-tools/cycle8/alma-proposers-guide}. These models
were initialized using the
stratospheric temperature profile retrieved from recent, unresolved ALMA
observations of Titan in 2019 \citep{thelen_20}, along with
tropospheric temperature profiles (which are not expected to show
significant seasonal variability)
from Cassini Radio Science
measurements \citep{schinder_20} interpolated to the corresponding
latitude regions. The spectral scaling factors were found to be between
0.85--1.15 for each spectral window.

\begin{figure}
  \centering
  \includegraphics[scale=0.8]{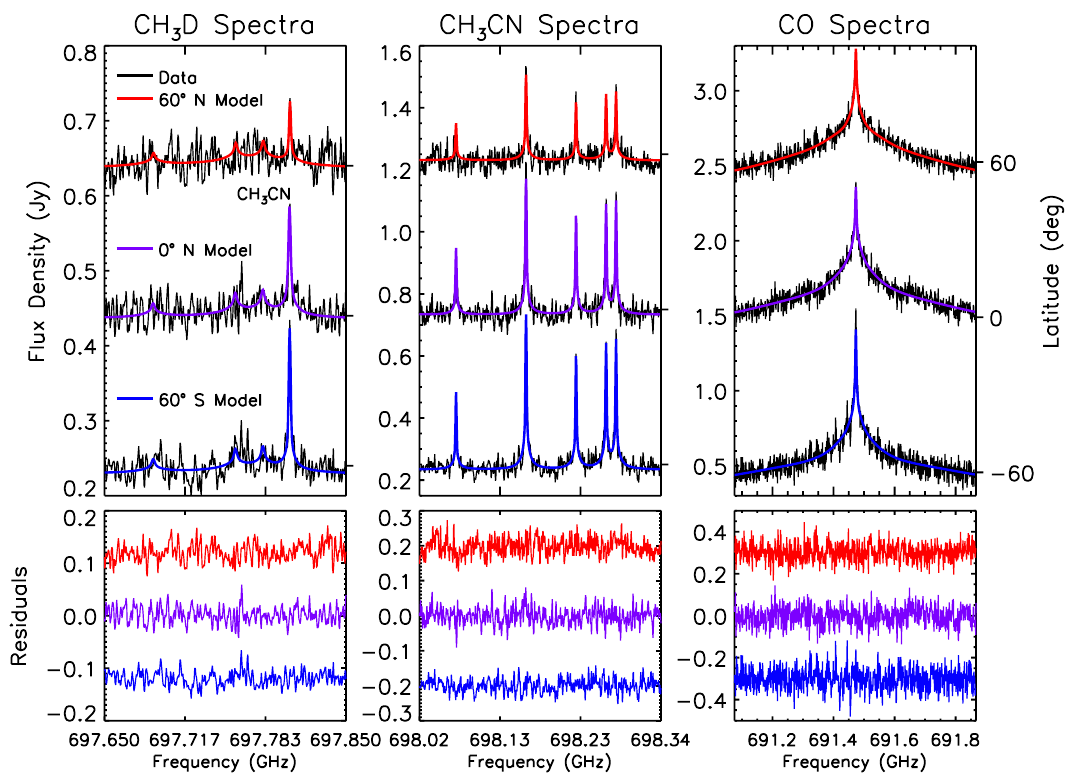}
  \caption{Top row: Representative ALMA data (black) compared to synthetic spectra
    from the corresponding best-fit NEMESIS models at 60$^\circ$N (red), the equator
    (purple), and 60$^\circ$S (blue) for spectral ranges covering the CH$_3$D (left column),
    CH$_3$CN (middle column), and CO (right column) transitions. Spectra are offset by arbitrary,
    constant factors for clarity (0.2 Jy for CH$_3$D; 0.5 Jy for CH$_3$CN
    and CO). The interloping CH$_3$CN ($J=38_6$--$37_6$) transition at
    697.804 GHz is denoted in the equatorial CH$_3$D spectrum. Bottom row: residual (data--model)
    spectra corresponding to the data and models in the top
    row. Residual spectra are offset for clarity (0.12 Jy for CH$_3$D;
    0.2 for CH$_3$CN; 0.3 for CO).}
  \label{fig:ret_spec}
\end{figure}

The temperature profiles between the stratosphere and mesosphere were
then retrieved using the CO ($J=6$--5) emission line. Retrievals at each latitude bin were 
performed by allowing the disk-averaged temperature profile from
\citet{thelen_20} to continuously vary with altitude above $\sim100$ km, while keeping the
tropospheric temperatures from \citet{schinder_20} and a nominal CO abundance of $\sim50$ parts-per-million
(ppm; $1\times10^{-6}$), as determined by a number of studies (e.g.,
\citealp{de_kok_07, teanby_10b, gurwell_11, de_bergh_12, rengel_14,
  serigano_16}), held constant. Atmospheric temperatures were varied
iteratively with the NEMESIS retrieval algorithm (based on the
Levenberg-Marquardt principle -- see, e.g., \citealp{press_92}) to minimize
differences between the data
and model spectra (determined by the ``cost function'') and
subsequently produced parameters that reached a global, reduced-$\chi^2$ minimum \citep{irwin_08}. The correlation length was taken
to be 1.5 scale heights, and though the profiles were allowed to vary
up to 1200 km, variations in the temperature profile above
$\sim600$ km ($1\times10^{-3}$ mbar) did not contribute significantly to minimizing the
reduced-$\chi^2$.
The resulting temperature profiles were then
used to retrieve the vertical CH$_3$CN abundance profile at each
latitude starting with a combination of
  abundance profiles from observations by \citet{marten_02} and
  photochemical predictions by \citet{loison_15} above $\sim400$ km,
  as was done with previous ALMA observations \citep{thelen_19a}. The
  strongest 5 spectral transitions of the CH$_3$CN
($J=38$--37) band were used to retrieve the vertical
profile without interference from the nearby CH$_3$D ($J=3$--2) triplet
(see Figure \ref{fig:spec}).

Finally, the aforementioned temperature and CH$_3$CN profiles were
fixed for CH$_3$D retrievals. The \textit{a priori} CH$_3$D
profile was generated from the constant stratospheric measurement of
CH$_4=1.48\%$ from the Huygens/GCMS 
\citep{niemann_10} and the D/H from the
weighted average of measurements from ALMA and Cassini/CIRS of
$1.2\times10^{-4}$ (see
\citealp{thelen_19b}, \citealp{nixon_12}, and references
therein). As the individual CH$_3$D spectra were relatively low
signal-to-noise, we performed retrievals on spectra from the East and
West hemispheres simultaneously to better constrain the CH$_3$D abundance. These models were parameterized as a simple scaling of
the input profile, as opposed to the continuous retrievals with altitude
that were employed for higher S/N spectral lines. Nearby CH$_3$CN lines
were included in the model, and were fit well by the input profiles
from the previous retrievals. Representative ALMA data for CO, CH$_3$CN, and
CH$_3$D are shown in Figure \ref{fig:ret_spec} compared to the
best-fit NEMESIS models. Residual spectra show that the spectral lines
are fit well using the methods described above.

\section{Results $\&$ Discussion} \label{sec:dis}

\subsection{Temperature Profiles} \label{sec:temps}
\begin{figure}
  \centering
  \includegraphics[scale=0.83]{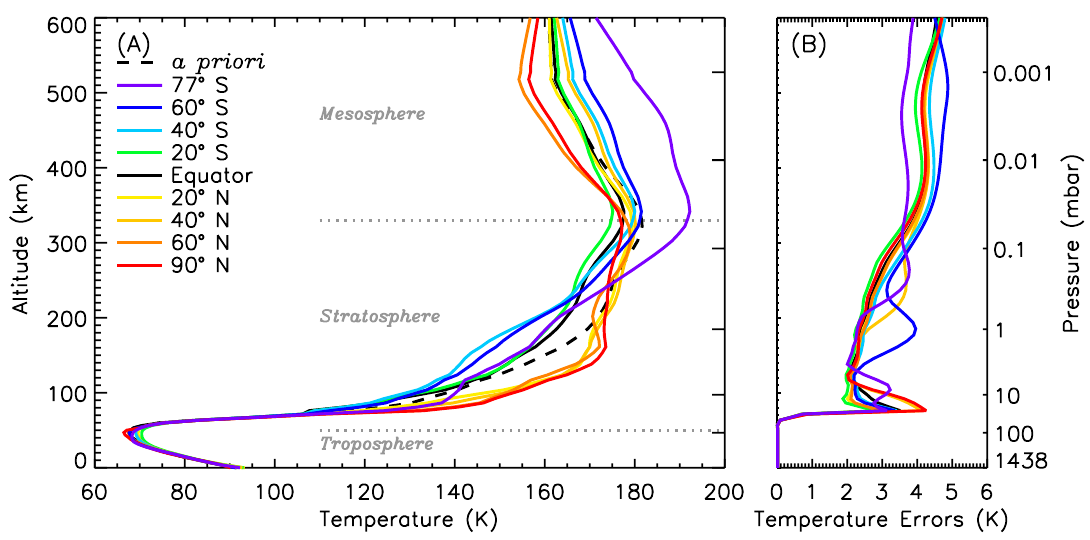}
  \includegraphics[scale=0.83]{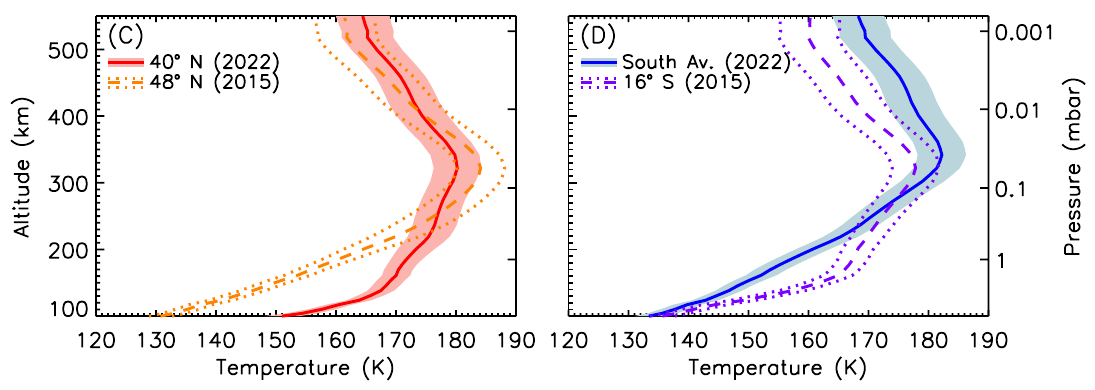}
  \caption{(A) The resulting latitudinally
    resolved temperature profiles (solid lines) retrieved using the
    ALMA CO ($J=6$--5) spectra. The input disk-averaged profile
    retrieved from previous ALMA observations \citep{thelen_20} is shown (dashed black line).
  (B) The retrieval errors corresponding to the vertical profiles in
  panel A. Errors below $\sim80$ km are tapered to 0 K to hold the
  tropospheric temperature values to those derived from the Cassini
  Radio Science data \citep{schinder_20}. (C) Comparison of the retrieved temperature profile from
    \citet{thelen_18} over a range of northern latitudes (centered at $\sim48^{\circ}$N; dashed
    orange profile and dotted errors) and the profile presented here
    at 40$^{\circ}$N (solid red profile and error envelope). (D)
    Comparison of the retrieved temperature profile from
    \citet{thelen_18} representing low southern latitudes ($\sim16^{\circ}$S; dashed
    purple profile and dotted errors) and
    an average of the profiles retrieved here from 20--77$^{\circ}$S 
    (solid blue profile and error envelope). The pressure axis on all panels is
    approximate, represented by the pressures from equatorial latitudes.}
  \label{fig:temp_comp}
\end{figure}

The resulting temperature profiles and corresponding errors retrieved from NEMESIS models of
the CO ($J=6$--5) transition are shown
in Figure \ref{fig:temp_comp}A and \ref{fig:temp_comp}B, respectively, throughout the stratosphere and mesosphere ($\sim100-500$ km; $\sim10-1\times10^{-3}$ mbar) 
where the spectral radiance is sensitive to changes in atmospheric temperature.
The retrieved temperature profiles show a distinct difference in regions north and south of the equator, particularly at higher
latitudes. Below the stratopause
(often near $\sim300$ km or 0.1 mbar), temperatures from the northern
hemisphere (warm colored profiles in Figure \ref{fig:temp_comp}A) are
elevated compared to those measured at both the equator (black profile) and southern
hemisphere (cool colored profiles) by $\sim10$--20 K,
particularly between 1 and 10 mbar ($\sim100$ to 200 km). Profiles
retrieved at 40$^\circ$ and 60$^\circ$S are cooler than the equator by
$\sim7$--10 K. Titan's radiative response timescale in the stratosphere is
short enough that the atmosphere will respond (i.e. equilibriate) to the increased insolation in the northern
hemisphere over $\sim1$ Earth yr, well within a Titan season
\citep{flasar_81}. Thus, the dichotomy between temperature profiles in
the northern and southern hemispheres follows naturally from Titan's
transition into its northern summer in 2017. While the
radiative response time is lower in the mesosphere -- so the altitudes
above 300 km respond rapidly to insolation -- the upper layer of Titan's
pole-to-pole circulation cell also affects these temperatures (see,
e.g., \citealp{teanby_08b, achterberg_11, teanby_12, vinatier_15}). We find
that profiles above 300 km ($\sim0.1$ mbar) are largely similar
$\pm40^\circ$ of the equator, but those at high latitudes show
influence of the meridional circulation. Northern hemispheric
profiles at high latitudes are cooler than those at lower latitudes above the
stratopause, likely due to air rising from below and cooling through adiabatic expansion. The subsiding branch of the
circulation cell increases temperatures at the high southern latitudes
and, in particular, the highest southern latitudinal profile we can
observe here (\textgreater77$^\circ$S) exhibits the highest and warmest stratopause
at 190 K between 320--350 km (purple profile in Figure
\ref{fig:temp_comp}A). While the profiles at low southern and
northern latitudes
are close together in magnitude, the stratopause may be
  slightly elevated ($\sim20$ km) at the highest southern latitudes
  (blue and purple lines in Figure \ref{fig:temp_comp}A) 
compared to the profiles at the highest northern latitudes (orange and
red lines). The measurement of temperature profiles in the
mesosphere and thermosphere can
be achieved using ALMA through the combination of CO and HCN emission
lines \citep{lellouch_19, thelen_22}, which we leave to future work
for these latitudes. 

\begin{figure}[h]
  \centering
  \includegraphics[scale=0.6]{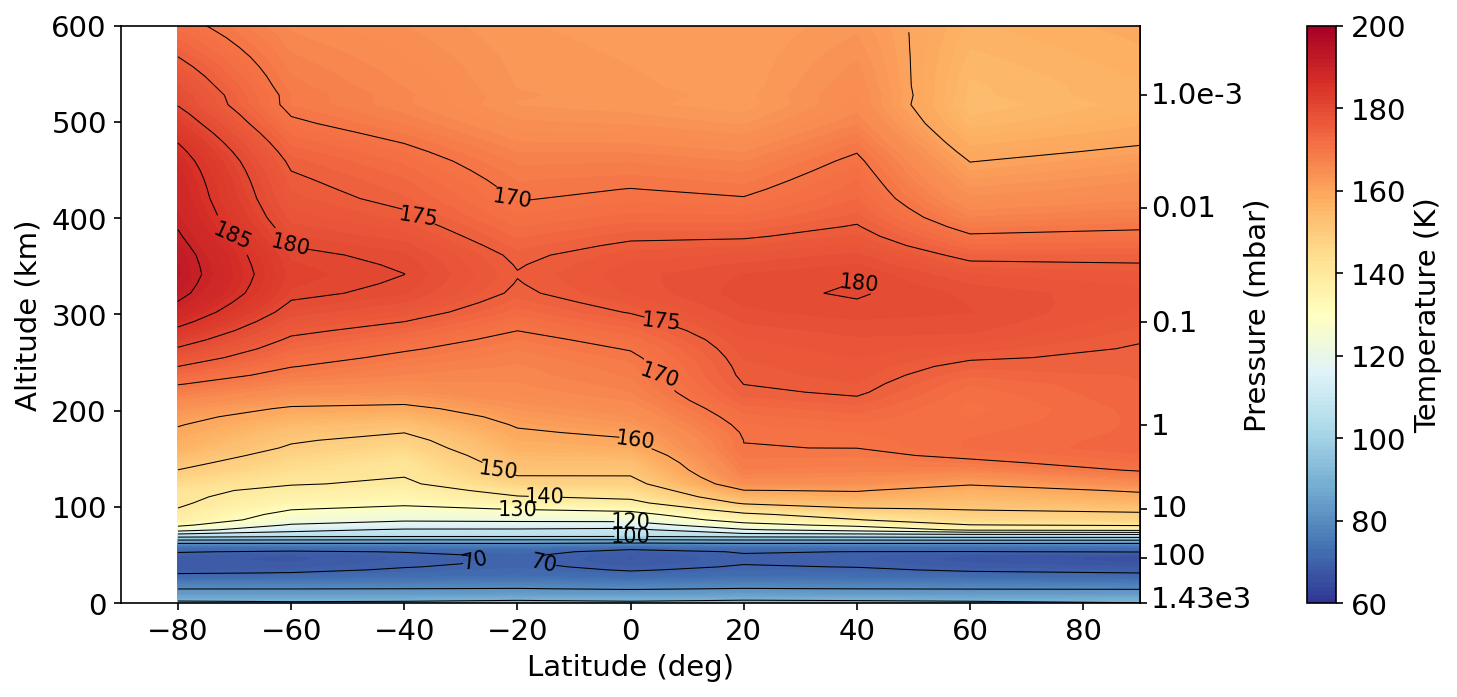}
  \caption{Temperatures from Figure \ref{fig:temp_comp}A, shown as a map as a
    function of latitude and altitude over the vertical range where
    our ALMA observations are sensitive. The latitudes below
    $80^\circ$ S are not shown due to Titan's sub-observer latitude at the
    time of these observations, which shrouds
    the south pole from view. Temperature contours are shown for
    reference, as is an approximate pressure grid from equatorial latitudes.}
  \label{fig:temp_map}
\end{figure}

We derive vertical temperature profiles that are similar to those
measured throughout the Cassini mission with the CIRS instrument
\citep{achterberg_08, teanby_10a, achterberg_11, vinatier_15, coustenis_18,
  teanby_19, coustenis_20, vinatier_20, mathe_20}, with lower
stratospheric (\textless200 km; \textgreater1 mbar) temperatures
ranging from $\sim130$--170 K and increasing to a stratopause
temperature of $\sim180$--200 K. We observe the continuation of the
trend observed through limb and nadir sounding of the atmosphere by
CIRS towards the end of the Cassini mission in an elevated, warmer southern stratopause, which
mirrors the initial measurements of Titan's northern polar
temperatures during its northern winter at the start of the Cassini
mission. This evolution is even apparent through the comparison of the
profiles presented here and ALMA observations from 2015 (Figure
\ref{fig:temp_comp}C and \ref{fig:temp_comp}D), which shows the increase in stratospheric
temperatures below $\sim200$ km in the north (Figure
\ref{fig:temp_comp}C), the cooling of the stratosphere at these
altitudes in the south, and the possibility of a slightly elevated and warmer southern
stratopause (Figure \ref{fig:temp_comp}D). It should be noted that the vertical and angular
  resolution of temperature profiles derived through these remote
  sensing observations is not sufficient to capture potentially stark latitudinal
  differences towards the poles as observed in other works (e.g. those from the Cassini
  limb sounding observations -- see, for example, \citealp{achterberg_08,
    vinatier_15}); as such, the magnitude of the stratopause altitude
  and temperature variability towards the poles may be somewhat diminished in these ALMA
  measurements. This is largely due to the ALMA beam size, which smoothed out
  the observed atmospheric state over $\sim10^\circ$ to $20^\circ$
  latitude ($\sim2$ to $5\times$ that of many CIRS observations), and
  did not allow for the high degree of altitude sampling as achieved
  with Cassini/CIRS limb observations \citep{kunde_96, flasar_04, nixon_19}. 

A decrease of northern stratopause temperatures was observed during
the course of the Cassini mission as Titan's northern hemisphere
transitioned out of winter and into spring, accompanied by the
elevation of southern stratopause temperatures through the adiabatic
heating induced by the subsiding branch of the meridional circulation cell -- particularly at
the south pole -- beginning shortly after Titan's 2009 spring equinox ($L_S=0^\circ$)
\citep{teanby_12, vinatier_15} and persisting until northern summer
solstice ($L_S=90^\circ$) in 2017 \citep{teanby_19, coustenis_18, coustenis_20,
  vinatier_20, mathe_20, achterberg_23}. The circulation of Titan's
middle atmosphere was predicted by modeling efforts and matched well
with Cassini observations (see, for example: \citealp{crespin_08, lora_15,
  shultis_22, lombardo_23a}). For comparison with these circulation
model studies, we
show our derived temperature profiles as a
function of latitude and altitude in Figure \ref{fig:temp_map}.
Though the viewing geometry of our observations prohibits the
retrieval of temperatures at the south pole, the temperature map shown
in Figure \ref{fig:temp_map} is in good agreement with the
distribution predicted by circulation models of Titan's northern
summer (see Figure 1 in \citealp{shultis_22}, or Figure 5 in \citealp{lombardo_23a}) and those found through IR
measurements towards the end of the Cassini mission
(see e.g., Figure 3 in \citealp{teanby_17}; Figure 2 in
\citealp{vinatier_20}; Figure 6 in \citealp{achterberg_23}); indeed, it is also similar to CIRS
measurements derived during Titan's northern winter, though now mirrored about the
equator \citep{achterberg_08, achterberg_11, vinatier_15}. Previous
observations also include a similar stratopause minimum at the low
latitudes in the winter hemisphere (see the saddle point at
$20^\circ$S in Figure \ref{fig:temp_map}). Future ALMA 
observations of Titan at similar angular resolution will allow for the
temporal measurement of the atmospheric temperature structure and assess changes over
Titan's northern summer and, in particular, during its autumnal
equinox ($L_S=180^\circ$) in 2025. Here, we may expect the large-scale circulation of
the middle atmosphere to break down and eventually reverse, thus
completing the seasonal cycle that began with the Cassini observations
of Titan in 2004. 

\subsection{CH$_3$CN Abundances} \label{sec:ch3cn}
\begin{figure}
  \centering
  \includegraphics[scale=0.83]{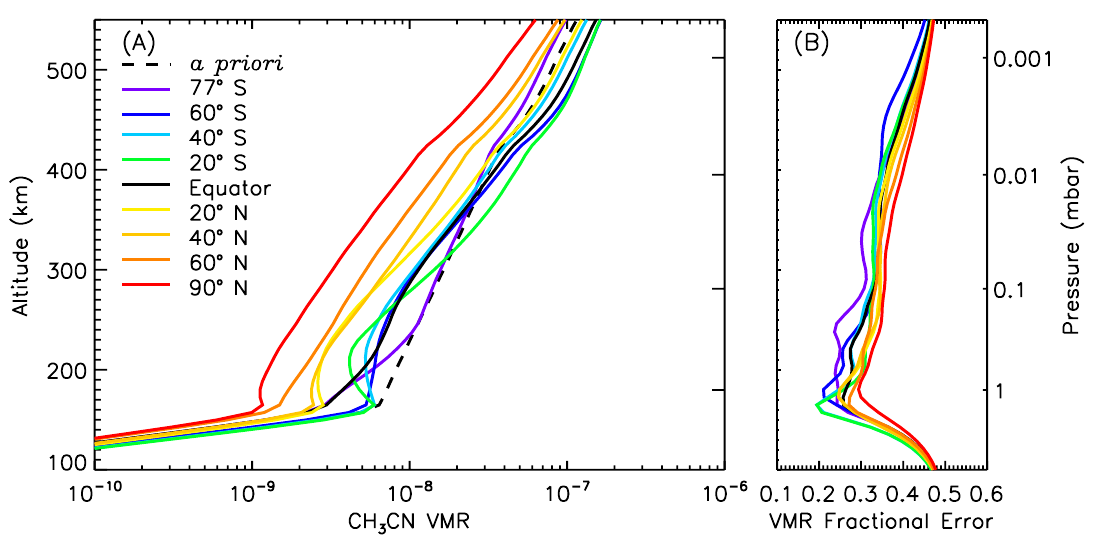}
  \includegraphics[scale=0.83]{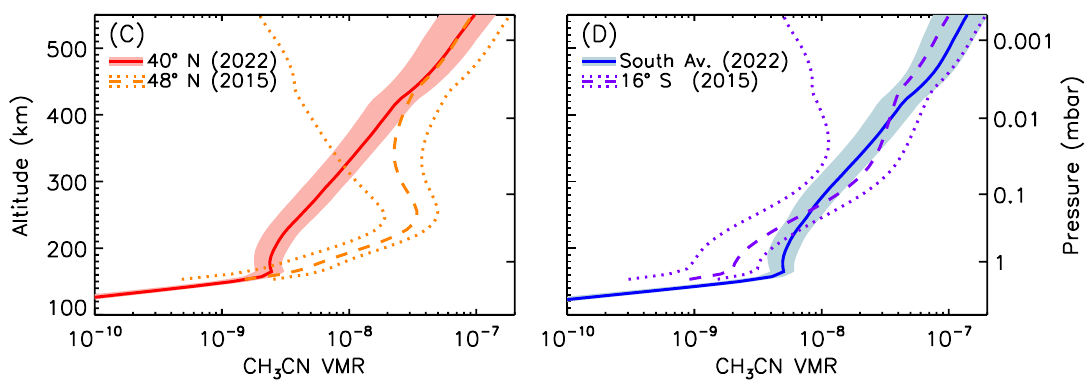}
  \caption{(A) The resulting latitudinally
    resolved profiles of CH$_3$CN abundance (solid lines) retrieved from
    ALMA CH$_3$CN ($J=38$--37) spectra. The retrievals are shown in 
    a similar altitude range as the retrieved temperature profiles
    (Figure \ref{fig:temp_comp}), though the
    CH$_3$CN spectra are not
    sensitive to abundance variations below $\sim160$ km and above
    $\sim450$ km. The \textit{a priori} CH$_3$CN profile, constructed
    from the observations by \citet{marten_02} and photochemical model
    predictions by \citet{loison_15} is shown (dashed black line). (B)
    The fractional VMR error profiles corresponding to the vertical
    profiles in panel A.
    (C) Temporal comparison of the CH$_3$CN vertical
  profile retrieved at $40^\circ$N (solid red line and error envelope) to ALMA observations
  in 2015 (dashed orange profile and dotted errors) from
  \citet{thelen_19a}. (D) Comparison of an average of
  southern hemisphere profiles derived here (solid blue line and error
  envelope) to the vertical profile representing low southern
  latitudes in 2015 (purple dashed profile and dotted errors) from
  \citet{thelen_19a}.}
  \label{fig:ch3cn_comp}
\end{figure}

The retrieved CH$_3$CN profiles are shown in Figure
\ref{fig:ch3cn_comp}A, and their errors (as a fraction of the
  CH$_3$CN abundance) in Figure \ref{fig:ch3cn_comp}B. As discussed in \citet{thelen_19a},
ALMA observations of CH$_3$CN are not sensitive
  to variations in abundance below $\sim160$ km; this is evidenced by the strong
influence of the chosen \textit{a priori} profile (here, taken from
\citealp{marten_02}) on the retrieved profiles at low
altitudes. 
The retrieved vertical profiles for CH$_3$CN show a decrease in volume
mixing ratio (VMR) with increasing northern
latitude, particularly at latitudes north of the equator. The equatorial profile
itself is generally in good agreement with the profiles derived by previous
(sub)millimeter observations \citep{marten_02, thelen_19a,
  lellouch_19} and photochemical model predictions above
\textgreater200 km \citep{loison_15, vuitton_19}. In the stratosphere
($\sim250$ km), the CH$_3$CN abundances at high southern latitudes are enhanced by a factor of
$\sim5$ compared to the north pole. At higher altitudes
(\textgreater400 km), where the
ALMA data become less sensitive, the abundance profiles are largely
similar to the equatorial profile south of $40^\circ$N. The differences in the vertical
profiles matches the emission map in Figure \ref{fig:maps}B, though
with some additional vertical structure at altitudes \textless200 km
potentially revealing the influence
of Titan's circulation cell on the CH$_3$CN distribution.

As the photochemical lifetime of CH$_3$CN in Titan's
middle atmosphere ($\sim300$ km) is predicted to be on the order of 
$1\times10^9$--$1\times10^{11}$ s ($\sim32$--3200 Earth yr; \citealp{wilson_04, loison_15, vuitton_19})
-- and thus significantly longer than Titan's $\sim29.5$ yr orbital
period -- the spatial distribution of CH$_3$CN will be driven
dynamically rather than by photodissociation and chemistry. The dynamical or transport lifetime of a species is
calculated for individual photochemical species based on vertical and
molecular diffusion \citep{loison_15,
  vuitton_19}, which can then be used to predict
the species' stratospheric polar enrichment when compared to its photochemical lifetime
and Titan's year \citep{teanby_09a, teanby_09b}. The observed distribution is
further influenced by vertical and horizontal
transport from Titan's winds and circulation. Indeed, the photochemical lifetime of CH$_3$CN is 
 longer than its dynamical lifetime by a factor
of $\sim100$ as calculated by the models of \citet{wilson_04},
\citet{loison_15}, and \citet{vuitton_19}, and comparable to that of
HCN.
Thus, it is reasonable to expect CH$_3$CN to be enriched by a factor of $\textless10$ at latitudes
\textgreater$60^{\circ}$N by comparison with the enhancements measured
by Cassini/CIRS during Titan's northern winter for HCN and other
long-lived molecular species,
which were found to be distinctly different than the detected
short-lived 
hydrocarbon species (e.g., C$_3$H$_4$, C$_4$H$_2$ ;\citealp{teanby_08a,
  teanby_09b, teanby_10b}). We measure an enhancement factor of $\sim1.5$ when comparing the profiles retrieved from spectra
at $\sim77^{\circ}$S (our furthest southern latitude) with that from
the equator (see Figure \ref{fig:ch3cn_comp}A). Despite the range of predicted
photochemical lifetimes in Titan's stratosphere for CH$_3$CN, our
factor is comparable to the predicted enhancement factor of 3 by
\citet{teanby_10b} based on other photochemical species and the
photochemical model of \citet{wilson_04}. However, without direct
measurements of CH$_3$CN at Titan's winter pole -- obscured by the
viewing geometry here -- it
is difficult to discern the exact enhancement factor. Our current
measurement places CH$_3$CN somewhat in line with 
the longer-lived (lifetimes \textgreater100 yr) hydrocarbon species (e.g., C$_2$H$_6$, C$_3$H$_8$) and
CO$_2$ as
observed during the Cassini mission, which were all observed to be
enhanced by factors of $\sim1-5$ \citep{teanby_08a,
  teanby_09a, teanby_09b, teanby_10b, coustenis_10}. This corroborates
the predicted dynamical lifetime of CH$_3$CN by \citet{loison_15} and
the transport lifetime by \citet{vuitton_19}, both of which are $\sim91$ yr at 300 km.

\begin{figure}
  \centering
  \includegraphics[scale=0.6]{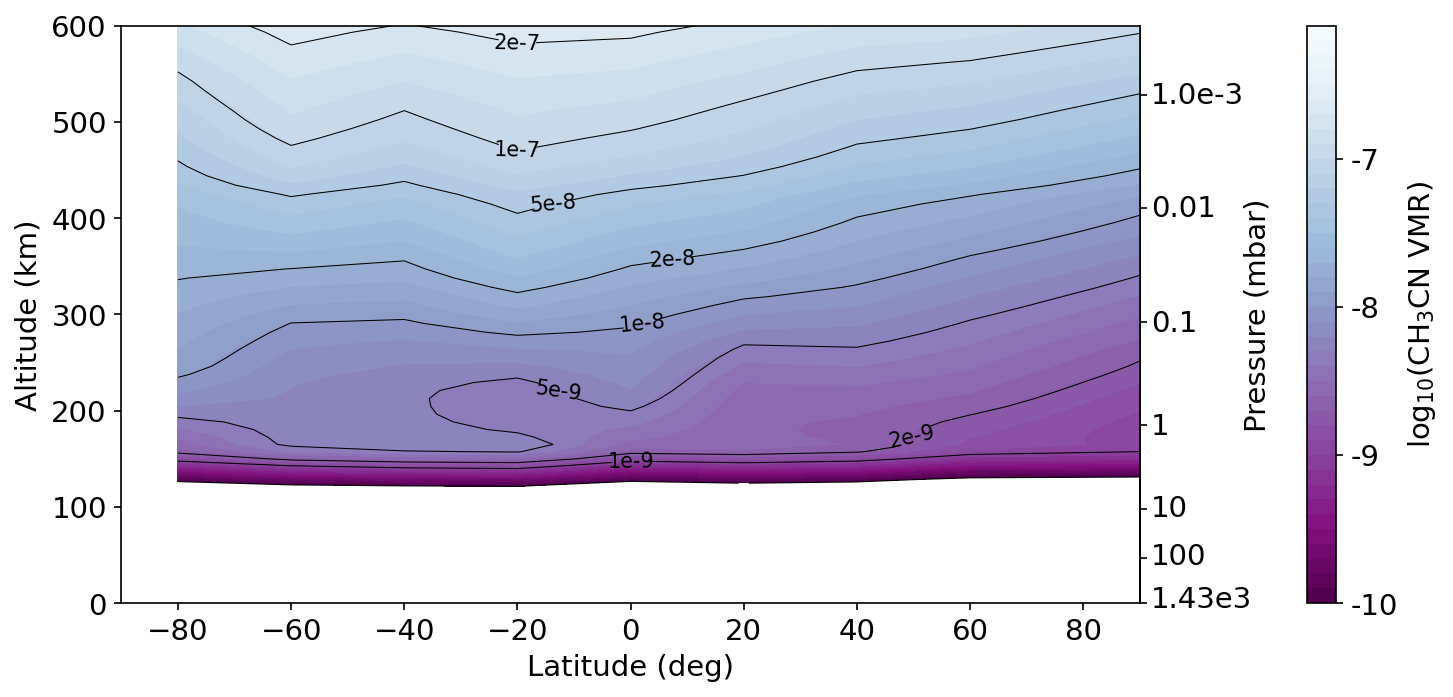}
  \caption{Map of CH$_3$CN abundances as a
    function of latitude and altitude. The pressure axis is approximate, represented by the pressures from equatorial latitudes.}
  \label{fig:chn_map}
\end{figure}

As with the temperature profiles, a comparison to previous ALMA
retrievals of CH$_3$CN abundance is shown in Figure
\ref{fig:ch3cn_comp}C and \ref{fig:ch3cn_comp}D. Here, we see a decrease in northern
abundances by a factor of $\sim16$ over roughly a Titan season (7
yr). Despite the stark enhancement in emission over the south pole shown in Figure \ref{fig:maps}B,
the abundances at high southern latitudes have only increased by a
factor of $\sim1.5-2$ over this time period. However, it should be
noted again that neither of these ALMA observations were able to
retrieve the true vertical profile over the south pole, where the
accumulation of CH$_3$CN may be more significant. Additionally, as
noted in past works \citep{thelen_19a}, the relatively large ALMA beam may preclude us
from measuring large enhancements with latitude due to smearing of the 
emission. Displaying the retrieved abundances as a
function of latitude and altitude (Figure \ref{fig:chn_map}) shows a
shallower CH$_3$CN abundance gradient than species with shorter
lifetimes as observed by Cassini/CIRS (e.g., HC$_3$N,
  C$_2$N$_2$; \citealp{teanby_09a, teanby_09b}),
but the increased abundances at altitudes \textless200 km between 20
and $60^\circ$S relative to the equator shows similarity to maps of the returning
circulation branch in the lower stratosphere, manifesting as a
`tongue' of enriched air moving equatorward \citep{teanby_08b}. This
was present in the similarly long-lived HCN during Titan's northern
winter, but surprisingly also in C$_4$H$_2$ \citep{teanby_08b}. Though
the predicted photochemical lifetimes of some species may be
incorrect, it is also possible that alternative or additional sources of
neutral and ion species production, such as galactic cosmic rays,
increase the stratospheric abundances \citep{loison_15}. Additional
observations of CH$_3$CN in the future may shed light on the changing
distribution of this species and the potential for it to track the
returning branch of the meridional circulation cell.

\subsection{CH$_3$D Abundances} \label{sec:ch3d}
\begin{figure}[h]
  \centering
  \includegraphics[scale=0.7]{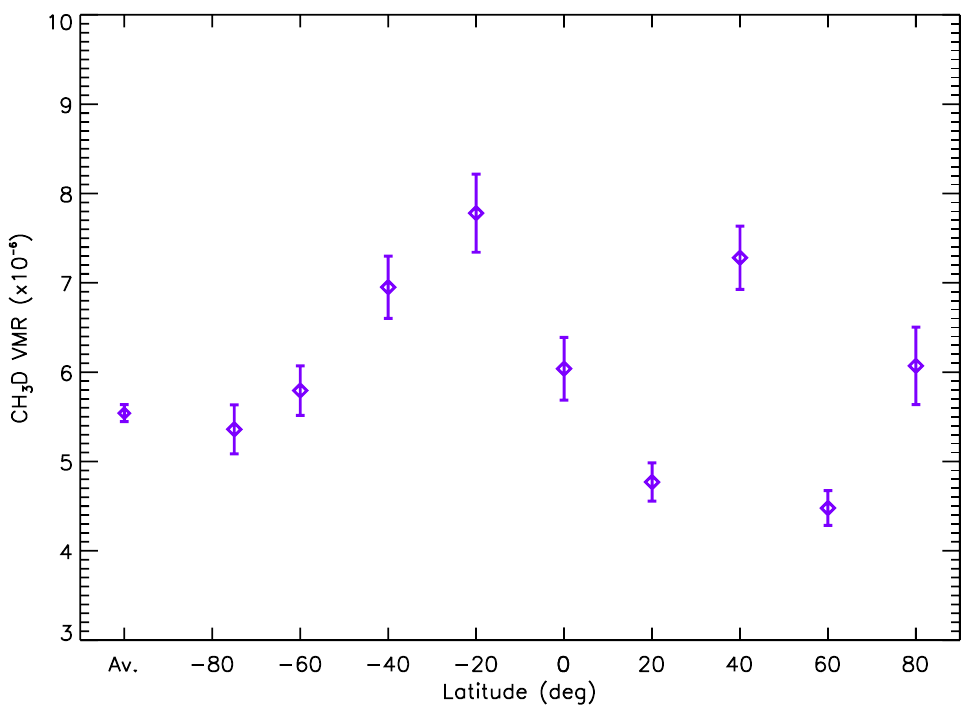}
  \caption{CH$_3$D abundance in Titan's stratosphere (between
    $\sim100$--300 km; $\sim0.1$--10 mbar). The retrieval error bars are
    shown for each data point. Each data point represents the averaged
    atmospheric CH$_3$D VMR for latitudes spanning $\sim\pm10^\circ$
    (low latitudes) to $\sim\pm20^\circ$ (high latitudes) due to the
    size of the ALMA beam and the sub-observer latitude of Titan ($\sim12^\circ$), as shown in
    Figure \ref{fig:maps}C. The weighted average of all
    observations is shown (left-most datapoint).}
  \label{fig:ch3d}
\end{figure}

\begin{figure}
  \centering
  \includegraphics[scale=0.7]{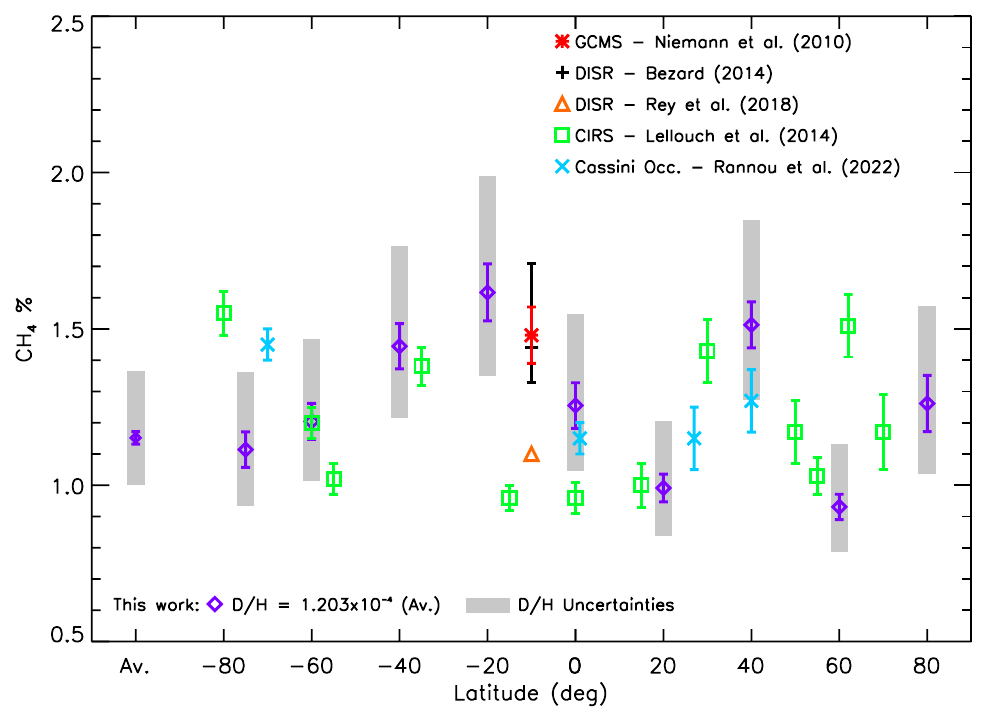}
  \caption{Converted CH$_4$ abundances as a function
    of latitude from Figure \ref{fig:ch3d} using
    the D/H derived in \citet{thelen_19b} are shown in purple. The weighted average
    of the converted CH$_4$ abundance is shown to the left. The shaded
    error envelopes illustrate the range of converted CH$_4$ abundances after
    considering the D/H ratios from ALMA (=$1.033\times10^{-4}$;
    \citealp{thelen_19b}) and CIRS (=$1.36\times10^{-4}$;
    \citealp{nixon_12} and references therein). The CH$_4$ abundances
    found by the Huygens GCMS \citep{niemann_10} and DISR
    \citep{bezard_14b, rey_18} are shown (red, black, and orange
    datapoints, respectively). Green squares show the
    average measurements of stratospheric
    CH$_4$ derived from Cassini/CIRS by
    \citet{lellouch_14}. Abundances between $\sim150-200$ km from
    Cassini occultation measurements are shown in blue crosses
    \citep{rannou_21, rannou_22}.}
  \label{fig:ch4}
\end{figure}

The stratospheric CH$_3$D abundances averaged
across both hemispheres are shown in Figure \ref{fig:ch3d}.
As in the emission map in Figure
\ref{fig:maps}A, there is significant spread in the retrieved CH$_3$D abundances. The range of
measurements, $\sim4-8$ ppm, encapsulates the values
previously measured by the ISO
\citep{coustenis_03}, ground-based observations \citep{penteado_05},
and Cassini \citep{bezard_07, coustenis_07, abbas_10, nixon_12,
  rannou_22}; the weighted average of all latitudes is
$(5.54\pm0.10)\times10^{-6}$ (shown as the left-most point in Figure
\ref{fig:ch3d}). When taken with the previously
derived D/H ratio based on the Cassini/Huygens CH$_4$ measurements, the converted CH$_4$ abundances are found to vary
between $0.9-1.6\%$, as shown in Figure \ref{fig:ch4}. We find that the values at $\pm40^\circ$ and $20^\circ$S
exhibit values around the $1.48\%$ level as measured by
the Huygens/GCMS \citep{niemann_10}. The converted weighted average
CH$_4$ abundance = $1.15\pm0.02\%$ (left-most point in Figure \ref{fig:ch4}), which agrees with
previous Cassini averages of $\sim1.1\%$ CH$_4$ in the stratosphere
\citep{lellouch_14, rannou_21, rannou_22}. These, and other
disk-averaged observations (e.g., those with the Herschel space
telescope -- \citealp{courtin_11, moreno_12a, rengel_14}),
indicate more variability in Titan's stratospheric methane than was
previously expected; many observations now indicate that the
stratospheric methane content is lower than the Huygens/GCMS
value derived during the probe's descent at $\sim10^\circ$S \citep{rey_18}. 

However, the range of CH$_4$ values we derive through CH$_3$D is
complicated by the lack of an
independent CH$_4$ profile, and thus the D/H uncertainty must be considered as well.
The global average CH$_4$ value in Figure \ref{fig:ch4} shows a
representative range of the inferred methane abundances when converted using
different D/H ratios, which range from $\sim(1.0-1.4)\times10^{-4}$ as
derived and discussed in \citet{nixon_12} and \citet{thelen_19b}. While the atmospheric D/H ratio is not expected
to vary with latitude, the range of measured D/H values adds
additional uncertainty to the CH$_4$ abundances inferred through the
measurement of CH$_3$D, increasing our range of converted CH$_4$ abundances to
$\sim0.8-1.9\%$ with a global weighted average of $1.15^{+0.21}_{-0.15}\%$.

The comparison of our converted CH$_4$ abundances to the variability
measured by \citet{lellouch_14} and \citet{rannou_22} with Cassini IR
and occultation measurements are also shown in Figure
\ref{fig:ch4} (green squares and blue crosses, respectively), as
are the abundances derived from the Huygens/GCMS and DISR instruments
\citep{niemann_10, bezard_14a, rey_18}. While our averaged and
equatorial 
measurements are lower than those measured by \citet{niemann_10} and
\citet{bezard_14a} with the Huygens probe, they are somewhat consistent
with the DISR re-analysis by \citet{rey_18} following the development
of newer spectroscopic CH$_4$ line data. In some instances, our converted
CH$_4$ abundances are in good agreement with those measured with
Cassini/CIRS and through VIMS occultations. At $40^\circ$S,
$20^\circ$S, and $40^\circ$N, however, the CH$_4$ values are elevated by $\sim0.4$--0.5$\%$ (i.e. 30--$40\%$ greater
than the global average value); these locations also correspond to the
enhancements on the integrated emission map (Figure \ref{fig:maps}A). \citet{lellouch_14} note that the
mixing timescale of CH$_4$ in Titan's stratosphere is long enough that
enhancements caused by tropospheric injection of methane-enriched air
may persist for a large portion of Titan's year; here, we see that
this may explain the agreement of values between these works at
certain latitudes, as the data analyzed by
\citet{lellouch_14} originate from Cassini observations almost 15 yr
prior to these ALMA observations. However, discrepancies where the
ALMA-derived abundances are considerably
higher (e.g., $20^\circ$S, $40^\circ$N) exist. These may be the result of more recent
injections that have altered the stratospheric CH$_4$ reservoir following
the Cassini measurements, as cloud activity was observed to be more
pronounced at 
mid-northern latitudes in Cassini/VIMS and Imaging Science Subsystem
data following Titan's spring equinox
\citep{turtle_18}. Additionally, the equinoctial
turnover of Titan's middle atmospheric circulation cell between these
observational epochs may be responsible for
redistributing locally enhanced
methane to southern latitudes; this may explain the difference between
the value retrieved here at $20^\circ$S and that found by
\citet{lellouch_19}. The dearth of CH$_4$ measured by ALMA in some
locations (e.g., 60$^\circ$N) compared to Cassini may be the result of the
previously observed enhancements relaxing to a well-mixed state due to
the changes in circulation during this period.

\citet{rannou_21} postulated that upwelling from more humid tropical
regions of the troposphere breaks through the tropopause and injects
higher abundances of CH$_4$ into specific, localized stratospheric
latitudes, which they find at $70^\circ$S, $\sim165$ km. Comparing the lower
stratospheric enhancement evident in the southern CH$_4$ vertical profile ($1.45\%$, as corrected in
\citealp{rannou_22}) to
equatorial measurements ($1.15\%$), abundances at 
higher stratospheric altitudes ($1.05\%$ at 250 km), and to the
measurements from 
\citet{lellouch_14} at a number of latitudes and altitudes, they propose a complex stratospheric CH$_4$
distribution influenced both by Titan's meridional circulation and
upwelling from localized humid regions of the troposphere based on
climate modeling predictions \citep{rannou_06} -- see Figure 3 in \citet{rannou_21}. As the Cassini CIRS
and VIMS observations were acquired during Titan's northern winter and spring,
we may expect both potential new enhancements in the stratosphere from
tropospheric injection in the
northern latitudes, as well as the redistribution of previously
observed enhancements throughout Titan's long seasonal cycle. Here, our measured enhancements at 40$^\circ$N and
80$^\circ$ N compared to the global average may be the result of
upwelling air at humid tropospheric latitudes between $\sim30$--$60^\circ$N,
which results in additional CH$_4$ distributed to higher northern latitudes and towards the
equator. A small, secondary circulation cell moving from mid-northern latitudes
poleward has been predicted by circulation models of Titan's atmosphere
during the northern summer
(see, e.g., \citealp{lombardo_23a}). We measure \textless$1\%$ CH$_4$
at $60^\circ$N, less than the previous measurement
of $\sim1.5\%$ in \citet{lellouch_14}, indicating that this may be primarily where
the tropospheric upwelling is occurring and bifurcating to
  poleward and equatorward circulation cells. Indeed, this is where the
polar circulation cell of \citet{lombardo_23a} is observed during northern
summer, mirrored from the southern summer circulation cell at
$\sim50$--$60^\circ$S in earlier models \citep{lora_15}. This is
invoked by \citet{rannou_21} to
explain their measured stratospheric methane distribution and the
relatively methane-dry value measured by \citet{lellouch_14} at $\sim55^\circ$S. Meanwhile the dropoff in
CH$_4$ from the
enhancement measured at $20^\circ$S to close to the average at
$80^\circ$S may be the result of the reversed meridional circulation
distributing CH$_4$ towards the southern hemisphere.

% Caveats:
Of course, these interpretations should be taken with caution, as the
distribution of CH$_4$ clearly appears to be complex, evolving,
and influenced by both middle and lower atmospheric circulation and
climate. Multiple factors contributing to these measurements and
interpretations are beyond the capabilities of our ALMA observations,
including: independent knowledge of the stratospheric methane itself, or the means
to derive the D/H separately from the reliance on previous measurements;
the temperature profile of the troposphere, which was taken into
account when discussing previous CH$_4$ measurements by
Cassini and used to infer the tropospheric methane abundance through
saturation vapor pressure for
comparison to stratospheric values. Currently, the relatively low S/N CH$_3$D emission also prohibits
the retrieval of a vertical profile, which would provide much more insight into
the influence of circulation on the interpretation of these
measurements (as it does for the other trace atmospheric
species). Local vertical enhancements of CH$_4$ were found at
different stratospheric altitudes compared to the background ($\sim1.1$--$1.2\%$)
in the
Cassini data analyzed by previous authors \citep{lellouch_14,
  rannou_21, rannou_22}, allowing for more complete inferences to be
made on the impact of the atmosphere's dynamical state on the CH$_4$
distribution. Our measurements of the CH$_4$ abundance should be taken as averages over a
large vertical range ($\sim100$--300 km); as such, the comparison of
these values to previous studies at both different times
and altitudes complicates an already inconsistent picture. The
proposed time-variable influences of
circulation, cloud activity, and tropospheric injection remain somewhat speculative until
further studies can improve both the D/H certainty and vertical
profile of the CH$_3$D measurements. Still, the aggregate CH$_3$D
measurements imply both a distribution of stratospheric methane that
is more complex than thought previously (corroborating the studies by 
\citealp{lellouch_14} and \citealp{rannou_21}), and that the global
average CH$_4$ abundance in the stratosphere -- even when including
the known spread of D/H values -- is depleted when compared to that of
the initial Huygens measurements. This indicates that the single
CH$_4$ profile from $\sim10^\circ$S in 2005 ($L_S\approx300^\circ$) should be used with caution when
considering other latitudes and epochs. 

\section{Conclusions} \label{sec:conc}
We have analyzed interferometric observations of Titan
in 2022 ($L_S\approx146^\circ$) following its northern summer solstice,
demonstrating seasonal changes in a heretofore sparsely investigated portion of
Titan's year. ALMA Band 9 observations ($\sim690-710$ GHz) enabled the retrieval of
abundance and temperature data from multiple latitudinal regions
across the disk, which allow us to continue monitoring the atmospheric
structure and chemistry after the end of the Cassini mission. Through
the analysis of these observations, we find:
\begin{itemize}
\item Titan's stratospheric temperature structure in 2022 is similar to that
  near the northern summer solstice in 2017, and mirrors the
  distribution of stratospheric temperatures observed during the northern
  winter at the start of the Cassini mission (2004; $L_S\approx293^\circ$).
\item For the first time, we retrieve vertical profiles of acetonitrile
  (CH$_3$CN) abundance that exhibit an enhancement in the
  southern hemisphere. Abundances are
  enhanced by factors $\sim1.5$--5 compared to the equator and the north
  pole, but the northern hemisphere abundances have been depleted by
  up to a factor of 16 since ALMA observations in 2015 (during
  northern spring).
\item Measurements of stratospheric CH$_3$D sensitive to averaged
  abundances between $\sim100$--300 km ($\sim0.1-10$ mbar) reveal a complex distribution of monodeuterated
  methane, which appears at odds with previous measurements in the IR
  and from the Huygens probe. Using previously derived D/H values, the
  inferred CH$_4$ content of the stratosphere can be found to vary
  between $\sim0.8-1.9\%$, with a global average of
  $1.15^{+0.21}_{-0.15}\%$.
\item Together, these temperature and abundance measurements are
  indicative of the influence of Titan's large, meridional
  circulation, which warms the upper stratosphere of the winter pole
  (currently, the south pole) and concentrates the long-lived trace
  atmospheric species inside the polar vortex. The distribution of methane is further impacted
  by Titan's tropospheric climate, which complicates the
  interpretation of the CH$_4$ distribution, but is in general agreement with
  previous models used to explain the stratospheric distribution
  during northern winter. 
\end{itemize}
  Future observations with ALMA will enable the continued measurement
  of Titan's atmospheric structure and composition, including into
  higher altitudes than were typically possible using Cassini infrared
  data for some chemical species \citep{thelen_22}. Higher sensitivity (sub)millimeter investigations
  of Titan's CH$_3$D may allow for the retrieval of vertical profiles,
  which will shed additional light onto the impacts of tropo- and
  stratospheric climate and circulation on Titan's stratospheric
  methane. 

\section{Acknowledgments}
Funding for this paper was provided by the NASA ROSES Solar System
Observations program for AET, CAN, and MAC. EL,
SV, and RM thank the French ``Programme National de Plan{\'e}tologie''
for funding.

The authors would like to acknowledge M. Palmer for their contribution
to the obeservation proposal. We would also like to acknowledge L. Barcos-Mu{\~n}oz,
R. Loomis, and the North American ALMA Science Center staff for their
knowledge and support during the imaging stages of the data reduction.

This paper makes use of the following ALMA data:
ADS/JAO.ALMA$\#$2021.1.01388.S. ALMA is a partnership of ESO
(representing its member states), NSF (USA) and NINS (Japan), together
with NRC (Canada), MOST and ASIAA (Taiwan), and KASI (Republic of
Korea), in cooperation with the Republic of Chile. The Joint ALMA
Observatory is operated by ESO, AUI/NRAO and NAOJ. The National Radio
Astronomy Observatory is a facility of the National Science Foundation
operated under cooperative agreement by Associated Universities, Inc.

\end{document}